\documentclass[letter,traditabstract,longauth]{aa}
\usepackage{graphicx}
\usepackage{txfonts}

\usepackage{natbib}
\newcommand{\jms}{J.~Mol.~Spectr.}                
\newcommand{\jpcrd}{J.~Phys.~Chem.~Ref.~Data}  

\begin{document}
   \title{Detection of anhydrous hydrochloric acid, HCl, in IRC+10216 with the
   $\it{Herschel}$ SPIRE and PACS spectrometers
   \thanks{$\it{Herschel}$ is an ESA space observatory with science instruments provided by European-led Principal Investigator consortia and with important participation from NASA}
   }
   \subtitle{Detection of HCl in IRC+10216}
   \authorrunning{J. Cernicharo et al.}
   \titlerunning{Detection of HCl in IRC+10216}
\author{
J.~Cernicharo\inst{1},
L.~Decin\inst{2,3},
M.J.~Barlow\inst{4},
M.~Ag\'undez\inst{1,5},
P.~Royer\inst{2},
B.~Vandenbussche\inst{2},
R.~Wesson\inst{4},
E.T.~Polehampton\inst{6,7},
E.~De Beck\inst{2},
J.A.D.L.~Blommaert\inst{2},
F.~Daniel\inst{1},
W.~De Meester\inst{2},
K.M.~Exter\inst{2},
H.~Feuchtgruber\inst{8},
W.K.~Gear\inst{9},
J.R.~Goicoechea\inst{1},
H.L.~Gomez\inst{9},
M.A.T.~Groenewegen\inst{10},
P.C.~Hargrave\inst{9},
R.~Huygen\inst{2},
P.~Imhof\inst{11},
R.J.~Ivison\inst{12},
C.~Jean\inst{2},
F.~Kerschbaum\inst{13},
S.J.~Leeks\inst{6},
T.L.~Lim\inst{6},
M.~Matsuura\inst{4,14},
G.~Olofsson\inst{15},
T.~Posch\inst{13},
S.~Regibo\inst{2},
G.~Savini\inst{4},
B.~Sibthorpe\inst{12},
B.M.~Swinyard\inst{6},
B.~Vandenbussche\inst{2},
C.~Waelkens\inst{2}
          }

\institute{
Departamento de Astrof\'{\i}sica, Centro de Astrobiolog\'ia, CSIC-INTA,
Ctra. de Torrej\'on a Ajalvir km 4, Torrej\'on de Ardoz, 28850 Madrid, Spain\\
\email{jcernicharo@cab.inta-csic.es}
\and
Instituut voor Sterrenkunde, Katholieke Universiteit Leuven,
Celestijnenlaan 200D, 3001 Leuven, Belgium
\and
Sterrenkundig Instituut Anton Pannekoek, University of
Amsterdam, Science Park 904, NL-1098 Amsterdam, The
Netherlands
\and
Dept of Physics \& Astronomy, University College London,
Gower St, London WC1E 6BT, UK
\and
LUTH, Observatoire de Paris-Meudon, 5 Place Jules Janssen, 92190 Meudon, France
\and
Space Science and Technology Department, Rutherford Appleton
Laboratory, Oxfordshire, OX11 0QX, UK
\and
Department of Physics, University of Lethbridge, Lethbridge,
Alberta, T1J 1B1, Canada
\and
Max-Planck-Institut f{\"u}r Extraterrestrische  Physik,
Giessenbachstrasse, 85748, Germany
\and
School of Physics and Astronomy, Cardiff University, 5 The
Parade, Cardiff, Wales CF24 3AA, UK
\and
Royal Observatory of Belgium, Ringlaan 3, B-1180 Brussels,
Belgium
\and
Blue Sky Spectroscopy, 9/740 4 Ave S, Lethbridge, Alberta T1J
0N9, Canada
\and
UK Astronomy Technology Centre, Royal Observatory
Edinburgh, Blackford Hill, Edinburgh EH9 3HJ, UK
\and
University of Vienna, Department of Astronomy, T{\"u}rken\-schanz\-stra\ss{}e 17, A-1180 Vienna, Austria
\and
Mullard Space Science Laboratory, University College London,
Holmbury St. Mary, Dorking, Surrey RH5 6NT, UK
\and
Dept of Astronomy, Stockholm University, AlbaNova University
Center, Roslagstullsbacken 21, 10691 Stockholm, Sweden
          }

   \date{Received / accepted}


  \abstract
   {We report on the detection of anhydrous hydrochloric acid (hydrogen chlorine, HCl) in the carbon-rich
    star IRC+10216 using the spectroscopic facilities onboard the $\it{Herschel}$ satellite.
    Lines from J=1-0 up to J=7-6 have been detected. From the observed intensities, we
    conclude that HCl is produced in the innermost layers of the circumstellar envelope
    with an abundance relative to H$_2$ of 5$\times$10$^{-8}$ and extends until the molecules reach its 
    photodissociation zone. Upper limits to the column densities of AlH, MgH,
    CaH, CuH, KH, NaH, FeH, and other diatomic hydrides have also been obtained.}

\keywords{Stars: individual: IRC +10216 --- stars: carbon
--- astrochemistry --- line:identification --- stars: AGB and post-AGB}

   \maketitle
%

\section{Introduction}

The chemistry of chlorine (Cl) in the interstellar medium (ISM) is
particularly poorly known, mainly because of the relatively small number 
of Cl-containing molecules detected to date (see \citealt{neu09}).
This element has two stable isotopes ($^{35}$Cl
and $^{37}$Cl) and has a relatively low solar abundance of
3$\times$10$^{-7}$, relative to H
\citep{asp09}. Anhydrous hydrochloric acid (HCl; also known
as hydrogen chloride) remains the only chlorine-bearing molecule 
observed to date in the interstellar
medium, and is believed to be one of the major reservoirs of
chlorine in the ISM. It has been observed in both the dense and
diffuse interstellar medium \citep{bla85,fed95}. The three metal
chlorides AlCl, NaCl, and KCl have also been observed in space
\citep{Cernicharo1987}, but solely in circumstellar envelopes (CSEs)
around
evolved stars, where they are formed in the hot and dense stellar
atmospheres under thermochemical equilibrium.
Although it has not yet been observed in such environments
HCl is also expected to be a major chlorine species in circumstellar 
envelopes. Being a light
hydride, its rotational transitions lie in the submillimeter
and far-infrared domain, which is difficult to observe
from the ground because of severe atmospheric absorption. 
Other light species, mostly metal-bearing hydrides such as AlH, FeH, MgH,
and CaH, are detected in sunspots and M-type stars \citep{Gizis1997, 
Wallace2001} and are also potentially present in the
innermost zones of carbon-rich circumstellar envelopes.

The infrared source IRC+10216 (CW Leo) is one of the brightest in the sky,
making it an ideal target to be observed with the $\it{Herschel}$
Space Observatory (Pilbratt et al. , 2010). Around 50\% of
the molecules observed in space have been detected towards
this object. Most of those molecules are heavy carbon chain
radicals \citep{Cernicharo1996a}, 
metal-bearing species \citep{Cernicharo1987}, and both 
diatomic and triatomic molecules
\citep{Cernicharo2000}. 
Its far-infrared spectrum, obtained with low spectral resolution with the Infrared Space Observatory (ISO),
was analyzed by \citet{Cernicharo1996b}
with the Infrared Space Observatory (ISO) with limited spectral resolution. The spectrometers
on board $\it{Herschel}$ \citep{Pilbratt2010} offer the possibility to search for light diatomic
hydrides with high sensitivity, thanks to the telescope's large collecting
area and the performances
of the instruments, and with high spectral resolution compared to ISO.
In this Letter,
we report on the first detection of HCl toward the circumstellar
envelope of the carbon-rich star IRC +10216, and discuss the
implications for the chemistry of chlorine in these astronomical
regions. We also present upper limits to the abundance of metal hydrides.

\section{Observations and data reduction}
The three instruments on board the $\it{Herschel}$ satellite \citep{Pilbratt2010} have
medium to high spectral resolution spectrometers. 
PACS and SPIRE spectroscopic observations were obtained in the
context of the guaranteed time Key programme "Mass-loss of Evolved
StarS" (Groenewegen et al., in prep). 
The PACS instrument, its
in-orbit performance and calibration, and its scientific
capabilities are described in \cite{Poglitsch2010}. The PACS
spectroscopic observations of IRC +10216 consist of full SED scans
between 52 and 210 $\mu$m obtained in a 3$\times$1 raster, i.e., a
pointing on the central object, and two additional ones located 30" on
each side of it. The observations were
performed on Nov 12 2009 (OD 182). The position angle was 110
degrees. The instrument mode was a non-standard version of the
chop-nod PACS-SED AOT, used with a large chopper throw (6'). A
description of that mode and of the data reduction process can be
found in \cite{Royer2010}. The estimated global uncertainty 
in the line fluxes is 50 \%. However, the relative calibration
is much better, hence it is possible to estimate the contribution 
of the most abundant species to the lines of HCl
by using adjacent lines of these species.
PACS and SPIRE photometry observations are presented in Ladjal et al.
(2010).

\begin{figure}
\includegraphics[angle=0,width=8.1cm]{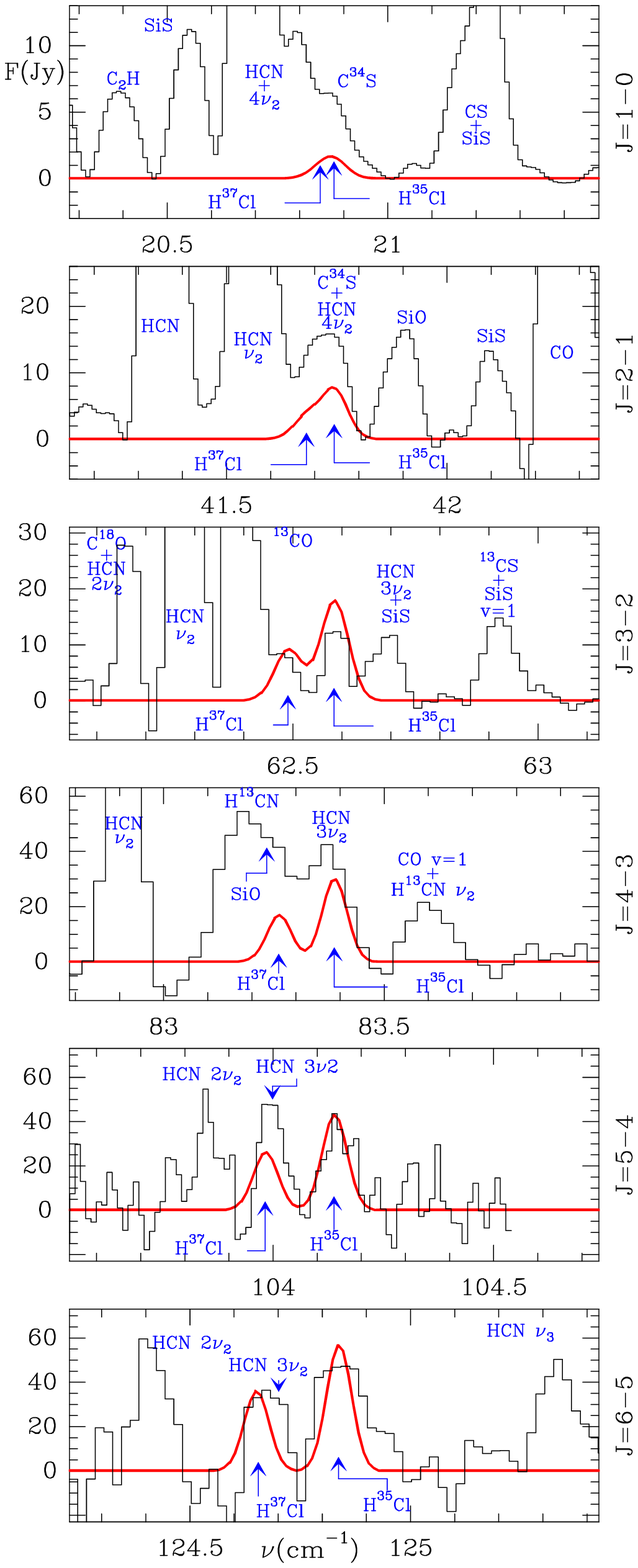}
\caption{Continuum-removed spectra of IRC +10216 observed with SPIRE and PACS
showing the first 6 rotational lines of HCl (black
histograms) and the line profiles resulting from the LVG model
(continuous red lines). The contribution of the isotopes
of CS and the vibrationally excited states of HCN to some
of the HCl and H$^{37}$Cl lines has been estimated from adjacent 
lines of these species (see text).} \label{fig-hcl-lines}
\vspace{-0.1cm}
\end{figure}

The SPIRE FTS measures the Fourier transform of the
source spectrum across short (SSW, 194–-313 $\mu$m) and long
(SLW, 303–-671$\mu$m) wavelength bands simultaneously. The
FWHM beamwidths of the central SSW and SLW pixels vary between
17-19" and 29-42", respectively. The source spectrum, including
the continuum, is restored by taking the inverse transform of the
observed interferogram. For more details about the SPIRE FTS and
its calibration, we refer to \cite{Griffin2010} and \cite{Swinyard2010}.
We made use of two observations of IRC+10216 with the high-resolution
mode of the SPIRE FTS on the 19 November 2009 (OD
189). For each observation, ten repetitions were carried out, each of which consisted
of one forward and one reverse scan of the FTS,
each scan taking 66.6s. The total on-source integration time for
each FTS spectrum of IRC +10216 was 1332 seconds. In the
end, both FTS spectra were averaged. The unapodized spectral
resolution is 1.2 GHz (0.04 cm$^{-1}$), which after apodization (using extended
Norton-Beer function 1.5; Naylor \& Tahic 2007) became
2.1 GHz (0.07 cm$^{-1}$). The sensitivity of the SPIRE/FTS spectrometer
allows us to detect lines as weak as 1-2 Jy. The whole PACS+SPIRE spectrum of 
IRC+10216 has been shown by Decin et al. (2010).

\section{Results}
In addition to lines arising from vibrational levels up to 
10000\,K \citep{Cernicharo1996b,Cernicharo2010}, HCN
is the main contributor to the spectral features detected with the SPIRE and PACS spectrometers. 
The frequencies of these lines were calculated by \cite{Cernicharo2010}
from the rotational constants provided by \cite{Maki1996,Maki2000}.
These frequencies were used to identify most features shown in Fig. 1.
The other strong features arise from CO, SiS, SiO, and CS. 
These lines were analyzed
by \cite{Decin2010}, and the data used to estimate
the contribution of these species to the lines of HCl shown in Fig. 1.
HCl has two stable isotopologs, 
H$^{35}$Cl and H$^{37}$Cl, the former being 3.1
times more abundant than the latter in IRC +102016, according to
the $^{35}$Cl/$^{37}$Cl abundance ratio derived from observations
of NaCl, KCl, and AlCl \citep{Cernicharo2000}. 
Frequencies for H$^{35}$Cl and H$^{37}$Cl were computed
from the rotational constants derived by \citet{caz04}. The
laboratory measurements have an accuracy better than 0.5 MHz for 
lines up to J$_{upper}$=14.

The first two rotational transitions of HCl are covered within the
spectral range of SPIRE, and are detected as emission lines in the
spectral data obtained towards IRC +10216 (see
Fig.~\ref{fig-hcl-lines}). 
The low spectral resolution of SPIRE
(2.1 GHz) prevents us from resolving the individual emission components 
related
to H$^{35}$Cl and H$^{37}$Cl, which are separated by 940 and 1879 MHz for
the $J$=1-0 and $J$=2-1 transitions, respectively. The $J$=1-0
emission is blended with the $J$=13-12 transition of C$^{34}$S.
Together, they appear as a shoulder at the high frequency side of
a stronger emission feature, which is a composite of several
rotational lines of HCN in different vibrational states
(we estimate a flux of 3 Jy for the J=1-0 line of HCl).
The $J$=2-1 transition is less severely blended
with stronger features, although it does overlap with the
$J$=26-25 transition of C$^{34}$S and with some $\ell$-doubling
components of the $J$=14-13 transition of HCN in its $\nu_2$=4
vibrational state. Inspection of the line intensities of
these species (C$^{34}$S and HCN $\nu_2$=4) in the nearby spectral
region (see Decin et al., 2010), indicates that they contribute at
most half of the intensity of the detected emission feature. Hence, the
measured flux for the $J$=2-1 transition of HCl is 9 Jy.

The spectral range of PACS covers the J=3-2 up to the J=9-8
rotational transitions of HCl. The data acquired
toward IRC +10216 allow one to clearly identify the $J$=3-2 to $J$=7-6
transitions (see Fig.~\ref{fig-hcl-lines}), the components related
to H$^{35}$Cl and H$^{37}$Cl being spectrally resolved. The
lines are clearly seen without the contamination of stronger lines
with derived fluxes, after removing the contribution from other species, 
of 12, 26, 40, 45, and 60 Jy for the J=3-2 to
J=7-6 lines, respectively.
Among them, only the $J$=4-3 rotational transition appears
appreciably contaminated, mostly by the $J$=29-28 transition of
H$^{13}$CN, which severely hampers the visualization of the
H$^{37}$Cl line, and to a lesser extent by the $J$=28-27
transition of HCN in its $\nu_2$=3 $\ell$=1 vibrational
level, which still leaves the H$^{35}$Cl line visible. 
The H$^{35}$Cl component of the $J$=7-6 transition
is observed as an emission
feature at the correct frequency and with an intensity compatible
with the lower $J$ transitions. However, the signal-to-noise ratio is only 5,
providing only upper limits to the intensity of the H$^{37}$Cl isotopomer.
Higher J lines are not detected because of the limited sensitivity at these
frequencies.
In spite of the low spectral resolution and the, in some cases quite severe,
contamination of the observed HCl lines, the large
number of transitions covered by the SPIRE and PACS data
makes the identification of HCl in the circumstellar gas of IRC+10216 
quite certain.

The observed HCl lines have been interpreted with the aid of an
excitation and radiative transfer model based on the Large
Velocity Gradient (LVG) formalism. The H$^{35}$Cl/H$^{37}$Cl
abundance ratio is poorly constrained from the observations and
was fixed to be 3.1, as derived for the $^{35}$Cl/$^{37}$Cl abundance
ratio from previous observations of NaCl, KCl, and AlCl in IRC
+10216 \citep{Cernicharo1987,Cernicharo2000}. We included the 
first 20 rotational levels
within the ground vibrational state of both H$^{35}$Cl and
H$^{37}$Cl. 
The adopted dipole moment is 1.109 D
\citep{del71}. The rate coefficients for collisional de-excitation
from HCl levels up to J=7, through collisions with H$_2$ and He,
were taken from
\citet{neu94}, and the Infinite Order Sudden (IOS) approximation
was applied for higher $J$ levels. The circumstellar envelope is
simulated as a spherically distributed gas expanding at a constant
velocity of 14.5 km s$^{-1}$. The gas density and temperature
radial profiles were taken from \citet{agu09} and \citet{Fonfria2008}. 
The adopted distance to IRC +10216 is 120 pc \citep{sch01}.

Molecules in IRC +10216 are either concentrated around the central
star (e.g., HCN) or distributed in a hollow shell of radius
10$''$-20$''$ (e.g. CN). The lack of information about the HCl
line profiles prevents one from deciding which of these two
distributions HCl follows. Several radial distributions for
HCl were tested, but the LVG model indicates that the observed
relative intensities of the HCl lines can only be reproduced if
HCl is concentrated around the central star. We thus
adopted an abundance radial profile in which the abundance of
HCl relative to H$_2$ is constant from the stellar photosphere
out to the radius where it is photodissociated by the
ambient interstellar UV field. The adopted photodissociation rate
for HCl is 1.1$\times$10$^{-9}$ $\exp$(-1.8 $A_V$) s$^{-1}$
\citep{rob91}, where $A_V$ is the visual extinction against
interstellar light at each radial position in the envelope. The
derived abundance of H$^{35}$Cl, relative to H$_2$, is
5$\times$10$^{-8}$, which produces line profiles that are in
reasonable agreement with the observed ones, as shown in
Fig.~\ref{fig-hcl-lines}.

\begin{figure}
\includegraphics[angle=0,width=8.9cm]{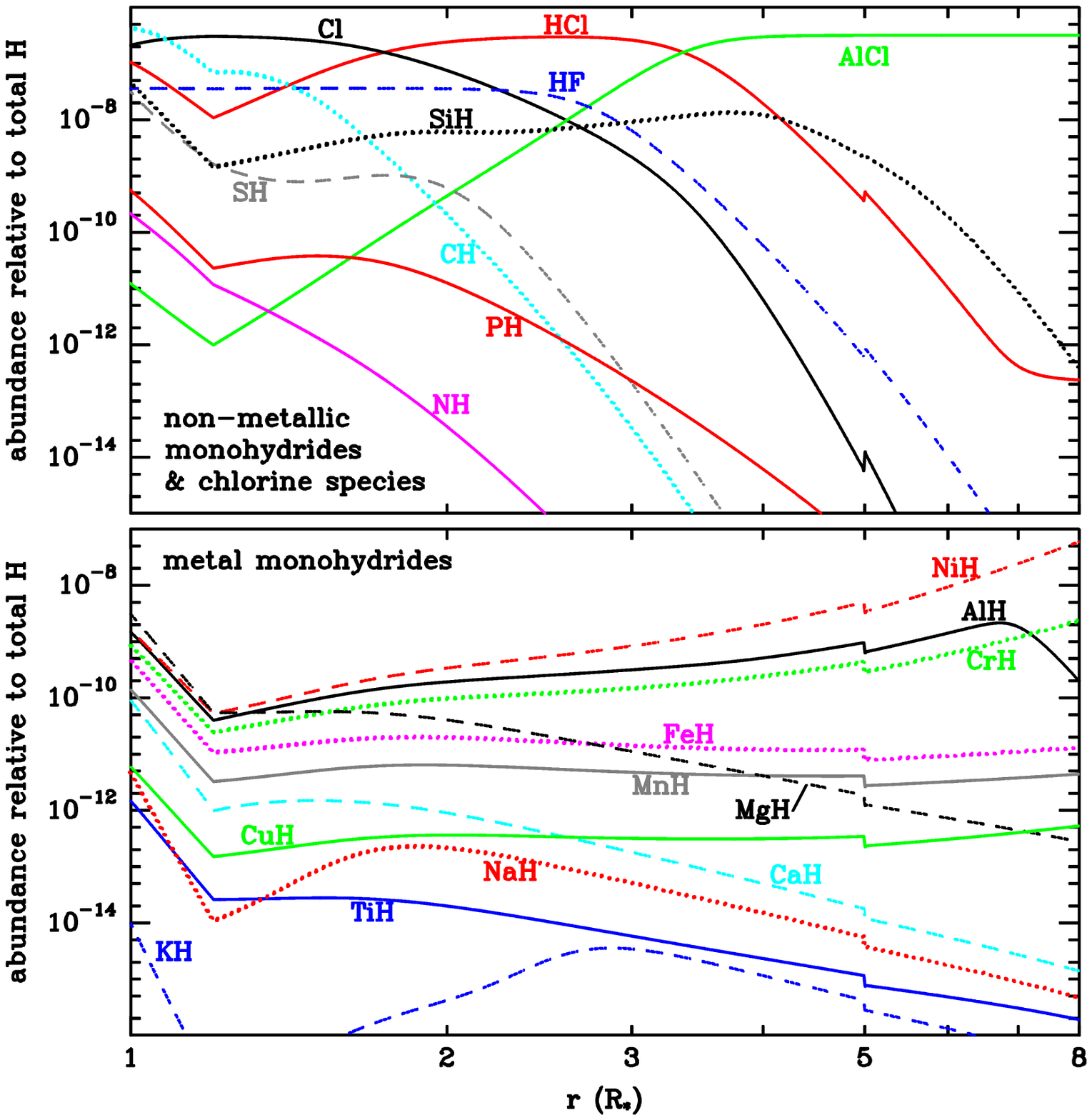}
\caption{Abundances of several hydrides computed by thermochemical
equilibrium for the stellar atmosphere and inner envelope of IRC
+10216. Abundances are relative to total H and are shown as a
function of the distance to the center of the star.}
\label{fig-abun-hydrides} \vspace{-0.1cm}
\end{figure}

\subsection{Chlorine chemistry in IRC +10216}
To verify whether the above
conclusion is compatible with chemical arguments, we
calculated the composition of the gas in the surroundings of the
stellar atmosphere of IRC +10216 under thermochemical equilibrium
(TE). The utilized code is described in \citet{tej91}. We
included 24 chemical elements with solar abundances \citep{asp09},
and assumed a higher carbon abundance so that
[C]/[O]=1.5. The thermochemical data of the included molecules
were taken from \citet{cha98}, with updates for various
species being taken from the recent literature. In particular, the
thermochemical data of TiH were updated according to \citet{bur05}.
The adopted parameters for IRC+10216
are presented in \citet{Agundez2006}. In Fig.~\ref{fig-abun-hydrides}, 
we 
show the calculated abundances
of several hydrides in the stellar atmosphere and inner envelope
of IRC +10216. We see that HCl reaches a maximum abundance of
6$\times$10$^{-7}$ relative to H$_2$ (i.e., half that value if
expressed relative to H) in the 2$-$3 R$_*$ region, where TE
still holds. According to the results shown in
Fig.~\ref{fig-abun-hydrides} the species that lock most of the
chlorine in the inner envelope of IRC +10216 are atomic Cl (in the
region inner to 2 R$_*$), HCl (in the 2$-$3 R$_*$ region), and
AlCl (at radii larger than 3 R$_*$). The HCl abundance derived
from the SPIRE and PACS data is lower than the calculated
thermochemical value by a factor of $\sim$10, which is 
acceptable given the uncertainties associated with the
observations and with both the LVG and the thermochemical
equilibrium models. Other Cl-bearing molecules observed in IRC
+10216 are AlCl, NaCl, and KCl \citep{Cernicharo1987}, for which the
abundances relative to H$_2$ derived from the IRAM 30-m data are
3.5$\times$10$^{-8}$, 1.0$\times$10$^{-9}$, and
2.5$\times$10$^{-10}$, respectively \citep{agu09}. Hence, HCl
and AlCl are the most abundant chlorine-bearing 
molecules in the inner envelope of
IRC +10216, HCl being slightly more abundant.
As the circumstellar gas expands, molecules become exposed to
the ambient UV field and are then
photodissociated. The chlorine carried by HCl will then be
liberated to the gas phase as atomic Cl, which in the outer
molecular shells (at 10$''$-20$''$) is just slowly ionized by
interstellar UV photons (the ionization potential of Cl is 13.0
eV, i.e., lower than, but very close to, that of hydrogen).
Therefore, it can participate in rapid chemical reactions to form new
Cl-bearing molecules. Which are the most likely such molecules is
difficult to predict because of the lack of
chemical kinetics data for reactions between atomic Cl and
abundant molecules in the outer envelope of IRC +10216.
The literature on the chemical kinetics of chlorine is vast, but
focuses mainly on species such as chlorofluorocarbons (CFCs),
which are of interest as important pollutants of the terrestrial
atmosphere. Some reactions of atomic Cl with unsaturated
hydrocarbons such as C$_2$H$_2$ have been also studied, although the
derived rate constant is normally for the three body process,
which is not of interest in the outer envelope of IRC +10216.
The UMIST Database for Astrochemistry \citep{woo07} contains
several reactions involving chlorine species, but is mostly oriented
toward the formation of HCl in interstellar clouds. 
Since atomic Cl is relatively reactive, it is expected to
react with abundant carbon-bearing molecules, such as the radicals
C$_2$H or C$_4$H, forming chlorine-carbon molecules. The detection
of these species (e.g., CCl, HCCl, C$_2$Cl, etc.) would certainly
aid in understanding the chemistry of chlorine in the outer layers
of circumstellar envelopes, which until to now has been a mystery.

\subsection{Other hydrides in IRC +10216}
As stated above, several metal hydrides have their rotational
transitions in the submillimeter and far-infrared domains.
These species are known to be abundant
in the photosphere of AGB stars \citep{Gizis1997}, and we expect
to detect them if they are not condensed onto dust grains.
We checked the rotational transitions of several of these
hydrides and we found no clear detection at the sensitivity limit of the data
(3 Jy for SPIRE and 10-20 Jy for PACS, depending on the
degree of blending, at the 3$\sigma$ level).
Moreover, many of their transitions are located in frequency domains
in which the line confusion limit is reached. This is mainly
the case
for the ranges associated with the HCN lines.
As an example, the J=5-4 and 10-9 lines of AlH
lie at 62.825 and 124.562 cm$^{-1}$, respectively. Neither line is
detected  (see Fig. 1). 
Assuming that the emitting region
for metal hydrides has an average kinetic temperature of 300 K, then the
3$\sigma$ upper limits to their abundance (relative to H$_2$)
are  3$\times$10$^{-9}$ (NaH), 9$\times$10$^{-9}$ (CuH), 
1.6$\times$10$^{-6}$ (AlH), 
4$\times$10$^{-8}$ ( MgH), 
3$\times$10$^{-7}$ (CaH), 
2$\times$10$^{-9}$ (LiH), 
3$\times$10$^{-9}$ (KH), and
10$^{-9}$ (FeH). We assumed that CaH and FeH have a dipole 
moment of one Debye.
In addition, SiH has its rotational
transitions in the SPIRE and PACS frequency domains. However, its dipole
moment is rather low and the upper limit to its abundance relative
to H$_2$ is 6$\times$10$^{-6}$.
Another interesting species is HF for which our
chemical models predict an abundance $\sim$10 times below that of HCl.
Unfortunately, its J=1-0 and J=2-1 lines are strongly blended with 
HCN vibrationally excited lines and those of other species.
The data used to estimate these upper limits are shown in
Fig. 1 of \citet{Decin2010}.

\section{Conclusions}
The detection of HCl in IRC+10216 indicates the active role of chlorine
in the chemistry of the warm innermost region of the circumstellar
envelope. Together with AlCl, HCl is the most abundant chlorine-bearing species in this
circumstellar envelope, in contrast NaCl and KCl having 
abundances 10 and 30 times lower, respectively. 
HIFI observations are
needed to spectroscopically resolve the rotational lines of light species
and to distinguish them from those of
more abundant molecules. These observations will provide more precise 
upper limits and might perhaps hold some detections of light species.

\begin{acknowledgements}
PACS has been developed by a consortium of institutes led by MPE
(Germany) and including UVIE (Austria); KUL, CSL, IMEC (Belgium);
CEA, LAM (France); MPIA (Germany); IFSI, OAP/AOT, OAA/CAISMI,
LENS, SISSA (Italy); IAC (Spain). This development has been
supported by the funding agencies BMVIT (Austria), ESA-PRODEX
(Belgium), CEA/CNES (France), DLR (Germany), ASI (Italy), and
CICT/MCT (Spain). SPIRE has been developed by a consortium of institutes led by
Cardiff Univ. (UK) and including Univ. Lethbridge (Canada);
NAOC (China); CEA, LAM (France); IFSI, Univ. Padua (Italy);
IAC (Spain); Stockholm Observatory (Sweden); Imperial College
London, RAL, UCL-MSSL, UKATC, Univ. Sussex (UK); Caltech, JPL,
NHSC, Univ. Colorado (USA). This development has been supported
by national funding agencies: CSA (Canada); NAOC (China); CEA,
CNES, CNRS (France); ASI (Italy); MCINN (Spain); SNSB (Sweden);
STFC (UK); and NASA (USA).
JC, MA, JRG, and FD thank spanish MICINN for funding
support under grants AYA2006-14876, AYA2009-07304 and CSD2009-00038.
LD and EDB
acknowledge financial support from the Fund for Scientific
Research - Flanders (FWO).
FK acknowledges funding by the Austrian Science Fund FWF under 
project numbers P18939-N16 and I163-N16.
JADLB, WDE, KME, RH, CJ, RR, and BV acknowledge support from the Belgian Federal 
Science Policy Office via the PRODEX Programme of ESA.

\end{acknowledgements}


\begin{thebibliography}{}

\bibitem[Ag\'undez \& Cernicharo (2006)]{Agundez2006}Ag\'undez, M., Cernicharo, J., 2006, \apj, 650, 374.
\bibitem[Ag\'undez (2009)]{agu09}Ag\'undez, M. 2009, PhD Thesis, Universidad Aut\'onoma de Madrid
\bibitem[Asplund et al. (2009)]{asp09}Asplund, M., Grevesse, N., Sauval, A. J., \& Scott, P. 2009, \araa, 47, 481
\bibitem[Blake et al. (1985)]{bla85}Blake, G. A., Keene, J., \& Phillips, T. G. 1985, \apj, 295, 501
\bibitem[Burrows et al. (2005)]{bur05}Burrows, A., et al. 2005, \apj, 624, 988
\bibitem[Cazzoli \& Puzzarini (2004)]{caz04}Cazzoli, G. \& Puzzarini, C. 2004, \jms, 226, 161
\bibitem[Cernicharo \& Gu\'elin (1987)]{Cernicharo1987}Cernicharo, J., \& Gu\'elin, M. 1987, \aap, 183, L10
\bibitem[Cernicharo \& Gu\'elin (1996a)]{Cernicharo1996a}Cernicharo, J., \& Gu\'elin, 1996a, \aap, 309, l27
\bibitem[Cernicharo et al. (1996b)]{Cernicharo1996b}Cernicharo, J., Barlow, M., Gonz\'alez-Alfonso, E., et al.,
1996b, \aaps, 315, L201
\bibitem[Cernicharo et al. (2000)]{Cernicharo2000}Cernicharo, J., Gu\'elin, M., \& Kahane, C. 2000, \aaps, 142, 181
\bibitem[Cernicharo et al. (2010)]{Cernicharo2010}Cernicharo, J., Ag\'undez, M., Kahane, C., et al., C. 2010, \apj, in press
\bibitem[Chase (1998)]{cha98}Chase, M., W. 1998, 
\jpcrd, Monograph n 9
\bibitem[Decin et al. (2010)]{Decin2010}Decin et al., 2010, \aap, this volume
\bibitem[De Leeuw \& Dymanus (1971)]{del71}De Leeuw, F. H. \& Dymanus, A. 1971, 26th Symposium on Molecular Spectroscopy, Columbus, Ohio
\bibitem[Federman et al. (1995)]{fed95}Federman, S. R., Cardelli, J. A., van Dishoeck, E. F., Lambert, D. L., \& Black, J. H. 
1995, \apj, 445, 325
\bibitem[Fonfr\'{\i}a et al. (2008)]{Fonfria2008}Fonfr\'{\i}a, P., Cernicharo, J., Ritcher, M.J., Lacy, J.H., 2008,
\apj, 673, 445
\bibitem[Gizis (1997)]{Gizis1997}Gizis, J.E., 1997, Astron. J., 113, 806-822
\bibitem[Griffin et al. (2010)]{Griffin2010}Griffin et al. 2010, \aap, this volume
\bibitem[Ladjal et al. (2010)]{Ladjal2010}Ladjal, D., et al. 2010, \aap, this volume
\bibitem[Maki et al. (1996)]{Maki1996}Maki, A., Quapp W., Klee, S., et al., 1996, J. Mol. Spectrosc., 180, 323
\bibitem[Maki et al. (2000)]{Maki2000}Maki, A., Mellau, G.Ch., Klee, S., et al., 2000, J. Mol. Spectrosc., 202, 67
\bibitem[Naylor \& Tahic (2007)]{Naylor2007}Nyalor, D.A., \& Tahic, M.K., 2007, J. Opt.
Soc. of America, 24, 3644
\bibitem[Neufeld \& Green (1994)]{neu94}Neufeld, D. A. \& Green, S. 1994, \apj, 432, 158
\bibitem[Neufeld \& Wolfire (2009)]{neu09}Neufeld, D. A. \& Wolfire, M. G. 2009, \apj, 706, 1594
\bibitem[Pilbratt et al. (2010)]{Pilbratt2010}Pilbratt et al. 2010, \aap, this volume
\bibitem[Poglitsch et al., (2010)]{Poglitsch2010}Poglitsch et al., 2010, \aap, this volume
\bibitem[Roberge et al. (1991)]{rob91}Roberge, W. G., Jones, D., Lepp, S., \& Dalgarno, A. 1991, \apjs, 77, 287
\bibitem[Royer et al., (2010)]{Royer2010}Royer et al., 2010, \aap, this volume
\bibitem[Sch\"{o}ier \& Olofsson (2001)]{sch01}Sch\"{o}ier, F. L. \& Olofsson, H. 2001, \aap, 368, 969
\bibitem[Swinyard et al., (2010)]{Swinyard2010}Swinyard et al., 2010, \aap, this volume
\bibitem[Tejero \& Cernicharo (1991)]{tej91}Tejero, J. \& Cernicharo, J. 1991, Instituto Geogr\'afico Nacional, Madrid
\bibitem[Wallace et al. (2001)]{Wallace2001}Wallace, L., Hinkle, K., \&  Livingston, W.C., 
2001. 
N.S.O. Technical Report  \#01-001, National Solar Observatory, Tucson
\bibitem[Woodall et al. (2007)]{woo07}Woodall, J., Ag\'undez, M., Markwick-Kemper, A., J., \& Millar, T., J. 2007, \aap, 466, 1197

\

\end{thebibliography}
\end{document}